\documentclass[useAMS,usenatbib,usegraphicx]{mn2e}

\newcommand{\CDC}{$^{12}$C/$^{13}$C~}
\newcommand{\Vt}{$V_{t}$}
\newcommand{\Teff}{T$_{\rm eff}$~}
\newcommand{\CO}{$\log(N_{\rm C}/N_{\rm O})$~}
\newcommand{\NO}{$\log(N_{\rm N}/N_{\rm O})$~}
	
\newcommand{\cm}{cm$^{-1}$}
\newcommand{\teff}{T$_{\rm eff}$}
\newcommand{\mim}{$\mu$m}

\newcommand{\JMS}{J. Mol. Spect.\ }
\newcommand{\JCP}{J. Chem. Phys.\ }

\newcommand{\AAA}{A\&A}
\newcommand{\AAAS}{A\&AS}
\newcommand{\ApJ}{ApJ}
\newcommand{\ApJS}{ApJSS}

\newcommand{\AJ}{AJ}

\title[HCN and HNC in Carbon Stars.]{The identification of HCN and HNC in Carbon Stars: Model Atmospheres, Synthetic Spectra and Fits to Observations in the 2.7-4.0 \mim\  Region}

\author[Harris et al.]{G. J. Harris$^1$, Ya. V. Pavlenko$^2$, H. R. A. Jones$^3$ and J. Tennyson$^1$\\
$^1$ Department of Physics and Astronomy, University College London, London, WC1E 6BT, UK.\\
$^2$ Main Astronomical Observatory, National Academy of Sciences, Zabolotnoho 27, Kyiv-127 03680, Ukraine.\\
$^3$ Astrophysics Research Institute, Liverpool John Moores University, Twelve Quays House, Egerton Wharf, Birkenhead. CH41 1LD, UK.}

\begin{document} 

\maketitle

%\altaffiltext{1}{}                               

%\altaffiltext{2}{Corresponding Author: }
                                         
\begin{abstract}

Model carbon star atmospheres and synthetic spectra have been calculated using the recent HCN/HNC vibration rotation linelist of \citealt{linepaper}, \ApJ, 578, 657. The calculations are repeated using only HCN lines and show that HNC has a significant effect upon the temperature, density and optical depth of a stellar atmosphere. 

We fit synthetic spectra in the 2.7-4.0 \mim\  region to observed ISO spectra of the carbon stars WZ~Cas and TX~Psc obtained by \citealt{Aoki1}, \AAA, 340, 222. These fits allow us to identify absorption by HNC in the spectrum of WZ~Cas at 2.8-2.9 \mim, and to determine new independent estimates of effective temperature and \CO.
The findings reported here indicate that absorption by both HCN {\it and} HNC is needed to fully explain the observed stellar spectra and represent the first identification of HNC in a star. Q branch absorption by the HCN $\Delta v_2=1$, $\Delta v_3=1$ and $\Delta v_1=1$, $\Delta v_2=-1$ bands at 3.55 and 3.86 \mim\  respectively, are identified in the spectrum of WZ~Cas.

\end{abstract}


\section{Introduction.}
\label{sec:Intro}

It is well known that HCN is an important opacity source in carbon stars and one which can dramatically affect the structure of the atmospheres of these stars \citep{Eriksson,Jorgensen1}. It is therefore vital that the opacity of this particular molecule should be accurately accounted for. Several recent works have used different sources of HCN opacity data.
The recent carbon rich model atmosphere calculations of \citet*{Aoki1,Aoki2} used the HCN band transition dipoles and molecular constants determined experimentally by \citet*{Maki95} and \citet{Maki96}, to account for HCN opacity.
 In contrast other recent works \citep*{Loidl,Loidl2001,Jorgensen2000} have used the ab initio data of \citet{Jorgensen1}.
Although ab initio data cannot rival laboratory data in accuracy, laboratory data are limited to tiny fraction of the billions of possible HCN transitions. As a consequence the laboratory data cannot possibly account for all of the HCN opacity. This makes ab initio data more useful for high temperature calculations which require a more complete description of molecular vibration rotation spectra and hence opacity.
The ab initio data of \citet{Jorgensen1} was the first extensive HCN opacity data set, and was a step forward from the empirical estimates of HCN opacity used by \citet{Eriksson}. 
However over the intervening decade and a half the advances in molecular quantum mechanical techniques coupled with increasingly more powerful computers has improved the quality and quantity of spectroscopic calculations on triatomic molecules.
These improvements enabled the calculation of the extensive HCN and HNC linelist of \citet*{linepaper}, which we use in this work. The linelist of \citet{linepaper} is both more extensive and accurate than the data of \citet{Jorgensen1}. 

A further advantage of the linelist of \citet{linepaper} over those of \citet{Jorgensen1} and \citet{Aoki1} is that it also takes into account the effect of the isomer of HCN: HNC. Recent calculations by \citet*{Bob} have yielded a HNC to HCN ratio, at thermodynamic equilibrium, of 0.13 at 2600 K and 0.19 at 3000 K. This coupled with the fact that the band transition dipoles of the fundamental bands of HNC are stronger than their HCN counterparts, imply that HNC will contribute to the opacity of C-star atmospheres.  

To determine the effect the new opacity data of \citet{linepaper} has on model C-star atmospheres and synthetic spectra, we present a grid of model atmospheres calculated using this data. We then discuss these model atmospheres and compare them with existing C-star model atmospheres. We compare the synthetic spectra with existing ISO observations of the C-stars TX Psc and WZ Cas in the 2.7 to 4 \mim\  region. This spectral region lies between the two strong CO bands, $\Delta$v=1 and 2, and contains several discernible HCN bands and the HNC H-N stretch fundamental band. From this comparison we determine effective temperatures and C/O ratios for these star. The effect of HNC opacity is identified for the first time in the spectrum of WZ Cas, and we attempt to identify Q branches of two HCN modes in the spectra of WZ Cas.

\section{HCN and HNC.}
\label{sec:sepHCNHNC}

The HCN/HNC linelist of \citet{linepaper} covers transitions between all HCN and HNC energy levels below 18~000 \cm\  above the HCN zero point energy and with angular momentum quantum number $J \leq$ 60. There are approaching 400 million lines and around 125~000 vibration rotation energy levels in this linelist. The accuracy of the results of this calculation room temperature situations has been discussed in \citet{spectpaper}. In particular it was found that the band transition dipoles calculated from this linelist were within combined systematic and experimental error. All the fundamental band centres are displaced from the laboratory values by less than 4 \cm, except the HNC H-N stretch fundamental which has an error of 12.5 \cm. The rotational error, corrected for vibrational error, for the lines of the fundamental vibrational modes of HCN/HNC are less than 0.7 \cm\  at J=20 \citep{surfpaper}.
Many of the lines in the linelist are extremely weak and make no real contribution to the overall opacity. We therefore reduce the number of lines used in the calculation of the model atmospheres and synthetic spectra by cutting out lines based on absolute intensity. We used a minimum absolute intensity at 3000 K of $10^{-15}$  cm$^{-2}$atm$^{-1}$, the strongest lines are of the order of 0.1 cm$^{-2}$atm$^{-1}$. This truncation dramatically reduced the number of lines, with little effect on the overall opacity. For the model atmospheres lines were also selected on the basis of proximity to one of the frequency points used in the calculation.
We must correct an error in the value of the constant given in equation 1 of \citet{linepaper}. This constant should read 3.56589$\times10^7$, and not 3.50729$\times10^{-6}$.

% separation of data.

To clearly show the effect of HNC opacity on the atmospheres of C-stars it is necessary to separate the combined HCN/HNC linelist of \citet{linepaper} into HCN and HNC data. There are 125~000 energy levels in the linelist of \citet{linepaper}, it would be very time consuming and tedious to identify by hand energy levels as HCN or HNC. Consequently we used the identification algorithm used by \citet*{spectpaper}. In this method individual energy levels are labelled HCN or HNC on the basis of the strengths of the transition from that level to the vibrational ground HCN and HNC states. An energy level that has a transition dipole to the ground state of HCN which is several orders of magnitude stronger than the transition to the ground HNC state, can confidently be identified as HCN and vice-versa. Two problems arise with this method. The first and most important involves energy levels which have very weak transitions to the ground state. For example the HCN (0,8,8,0)$\rightarrow$(0,0,0,0) band is forbidden by the normal dipole transition rules for a linear triatomic ($\Delta l=-1,0,+1$). This transition is so weak that it is below the noise level in the calculation, states with a transition to the ground state this weak we do not assign as HCN or HNC. Secondly, the idea that HCN and HNC are separate molecules is incorrect at high energies. The wave functions of some high lying states have magnitude in both HCN and HNC potential wells and so cannot be described as HCN or HNC, these are called `delocalised' states. Such states can have transitions to the HCN and HNC ground states which are of roughly the same order of magnitude. 

\begin{figure*}
%\epsscale{0.9}
%\plotone{HCN_abs.ps}
%\includegraphics{HCN_abs.eps}
\includegraphics{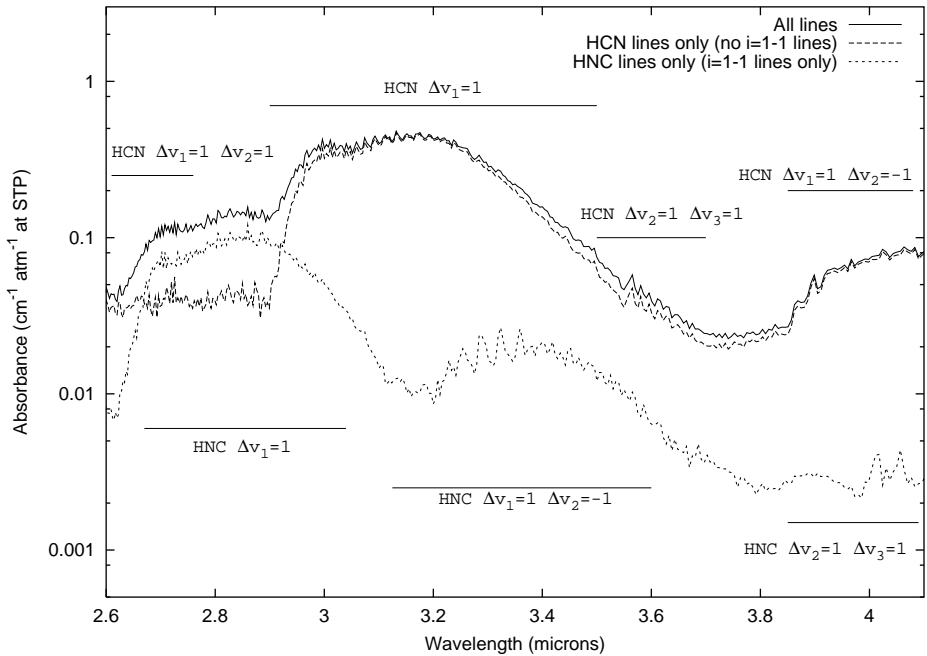}
\caption{The wavelength dependent absorbance of HCN and HNC. The three curves represent the absorbance of HCN/HNC (all lines), the absorbance of HCN (all lines except ($i^\prime=1$, $i^{\prime\prime}=1$) and HNC (lines with $i^\prime=1$, $i^{\prime\prime}=1$). The main HCN and HNC features are labelled. See text for a full description of $i$. }
\label{fig:HCN_abs}
\end{figure*}

The calculations that we make in this work which are described as `HCN only', exclude only the HNC levels which we can confidently identify. There may be other HNC energy levels in the data which are unassigned. However this `HCN only' data will exclude the strongest HNC transitions and will allow us to identify the effect of HNC on our model atmospheres and spectra. We label energy levels confidently identified as HCN as $i$=0, HNC levels as $i$=1, delocalised states as $i=2$ and unassigned states as $i=3$. Figure \ref{fig:HCN_abs} shows a plot of HCN and HNC absorbance at T=3000 K as a function of wavelength, from 2.6 to 4.1 \mim.
 The three curves show absorbance calculated with all lines, absorbance calculated with HCN lines only and absorbance calculated only with confidently identified HNC lines. 
Over this spectral range HNC absorption only exceeds that of HCN at wavelengths between 2.7 to 2.9 \mim. However between 3.2 and 3.9 \mim\ HNC appears to contribute significantly (10-15\%) to the total opacity. The effect of this HNC absorption can be seen in our synthetic spectra, see sections \ref{sec:WZCas_fit} and \ref{sec:TXPsc_fit}.

% fig1, HCN_abs

We have used the recent ab initio HCN/HNC partition function of \citet{Bob} in all of our calculations. This partition function was calculated by direct summation using the HCN/HNC energy levels of \citet{linepaper} up to J=32 and includes an extrapolation of these levels up to J=91, which is enough to rotationally converge the partition function. The error of this partition function has been estimated to be 0.6\% at T=2600 K and of 0.8\% at T=3000 K \citep{Bob}, HNC energy levels are also included in the calculation.

\section{Stellar Spectra.}
\label{sec:Star}

The spectral region required for this program is difficult to obtain from
ground-based observatories and is well matched to ISO capabilities. In
particular the SWS instrument on ISO provides sufficient resolution 
to resolve individual vibration rotation bands. 
The observations used for this study were taken as part of the 
Japanese guaranteed 
observing time (REDSTAR1, PI T.Tsuji, e.g., \citet{Aoki1}). 
This programme obtained 
a number of infrared measurements of Carbon stars. For the purposes
of our study we selected the high resolution observations made for
TX~Psc (N-type) and WZ~Cas (SC-type) from the ISO archive facility.

The observations were made in with the Short Wavelength Spectrometer
(SWS) in its full resolution grating mode known as SWS06. Here we 
use spectra with a resolution of 1000--2000
from bands 1B, 1D and 1E covering the 
wavelength ranges 2.74--3.02, 3.02--3.52 and 3.52--4.08 $\mu$m. 
The observational data are scientifically validated and 
were processed using the 
standard archive reduction software. The spectra were inspected 
using the 
ISO Spectral Analysis Package to check for glitches and bad data points.
\citet{Aoki1} note that the flux discrepancies are at most 
10\% for the REDSTAR1 dataset. For the subset of data used for 
this paper we find the flux calibration errors between SWS bands
are at the level of a few percent, as expected for band 1 SWS 
observations.
The wavelength calibration for the SWS
instrument is measured to be within 1/8 of a resolution element
\citet{Salama2000}. 

\section{Procedure}

\subsection{Ionisation-dissociation equilibrium}
In atmospheres of cool (\Teff $<$ 6500 K) carbon rich stars
the ionisation-dissociation equilibrium is governed by the CO molecule, 
that is, the  
densities of carbon-containing species depend on the 
C/O ratio (see Tsuji 1973 for more details). 
The ionisation-dissociation equilibrium is computed for
$\sim$ 100 of species including the most abundant carbon-containing molecules.
Constants of chemical equilibrium for most species are taken from \citet{Tsuji1973} and constants for HCN are 
converted into our model atmosphere format using partition function 
data of \citet{Bob} following the
scheme of \citet{Pavlenko2002b}. HCN and HNC molecules are treated as 
a single molecule with a single partition function and dissociation potential.  

\subsection{Opacities}

The standard set of continuum opacities included in ATLAS12 by \citet{Kurucz1999} are used. Computations of a few opacities 
were added or upgraded: the absorption of C$^-$ \citep{Myerscough} and H$_{2}^-$ \citep{Doyle}.
To compute
opacities due to bound-free absorption of C, N, O atoms we used TOPBASE
\citep*{Seaton}, (cross-sections are available on http://www.mao.kiev.ua/staff/yp/), see \citep{Pavlenko2003b} for more details.

To compute opacity due to bound-bound absorption of atoms and molecules 
in atmospheres of C-giants we used linelists from 
different sources:

\begin{enumerate} 

\item Electronic and vibration rotation line data of diatomic molecules (CN, C$_2$, MgH, H$_2$, SiH, CO) 
were  
taken from CDROM 18 of \citet{Kurucz1993}. For synthetic spectra computations 
we used the vibration rotation lines of CO from \citet{Goorvich1994}.

\item Atomic line data were taken from VALD \citep{Kupka1999}.

\item HCN/HNC vibration rotation lines were taken from \citet{linepaper}, see section \ref{sec:sepHCNHNC}. 

\item CS vibration rotation lines were taken from \citet{Chandra}.

\end{enumerate}

Other works \citep{Aoki1,Aoki2,Jorgensen2000}, have included C$_2$H$_2$ opacity. For warmer C-stars, such as WZ~Cas and TX~Psc, C$_2$H$_2$ has only a small effect on opacity \citep{Aoki1}. Hence it is not necessary to include H$_2$C$_2$ within our calculations. However for cooler C-stars with high \CO, H$_2$C$_2$ may well be more important and should be included within the calculation.

For the model atmosphere calculations we used the opacity sampling approach \citet*{Sneden1976} to account for the absorption by atomic and molecular lines.
Absorption of molecular electronic bands:
CaO (C$^1\Sigma$ - X$^1\Sigma$),  CS(A$^1\Sigma$ - X$^1\Sigma$),
SO (A$^3\Pi$ -  X$^3\Sigma$), SiO (E$^1\Sigma$-X$^1\Sigma$), 
SiO(A$^1\Pi$ -  X$^1\Sigma^+$, NO (C$_2 \Pi_r$- X$_2\Pi_r$),
NO(B$_2\Pi_r$ -X$_2\Pi_r$),
NO(A$^2\Sigma^+$ - X$_2\Pi_r$),
MgO(B$^1\Sigma^+$ -  X$^1\Sigma^+$),
AlO(C$^2\Pi$ -X$^2\Sigma$),
AlO(B$^2\Sigma^+$ - X$^2\Sigma^+$) are taken into account by the  
just overlapping line approach (JOLA), using the subprograms of \citet*{Ners1989}. In
general, these molecules absorb in the blue part of the spectrum and their
bands are not principal opacity sources in stellar atmospheres.
Nevertheless, the JOLA algorithm enables us to reproduce
both the relative strength and shapes of the spectral energy distributions
for late type stars \citep{Pavlenko1997,Pavlenko2000}.

%\subsection{linelists of atoms and molecules}

\subsection{Model atmospheres}
We computed a grid of model atmospheres of C-giants with the SAM12 program 
\citet{Pavlenko2002a,Pavlenko2003}
which is 
a modification of ATLAS12 (Kurucz 1999). SAM12 is a classical 
model atmospheres code: plane-parallel media, LTE, no 
sources and sinks of energy within the atmosphere, energy transferred by 
radiative field and convection. Convection is treated in the frame of 
mixing length theory, in our computations we adopt $l/H$ = 2.0, convective overshooting \citep{Kurucz1999} was not considered. 
Opacity
sampling treatment \citep{Sneden1976} is used to account for
atomic and 
molecular absorption. We investigate the sensitivity of our 
synthetic spectra to changes in effective 
temperatures \Teff, abundances of carbon log N(C) 
and microturbulent velocities $V_t$. 
We adopt the ``solar values'' of \citet{Anders} for 
elements other than carbon, for example oxygen log N(O) = -3.12 is 
adopted where $\sum N_i= 1$. 

Model atmospheres were computed for 
effective temperatures (\Teff = 2500 - 3500 K), carbon abundances 
log N(C) = -3.19 to -3.09, carbon isotopic ratio \CDC = 3 --10. Around 
100 model atmospheres were generated for this work.
The models computed with the complete HCN/HNC linelist are  
available from 
http://www.mao.kiev.ua/staff/yp/Results, as files Mod.Cg10.tar.gz
and Mod.Cg03.tar.gz.

\subsection{Synthetic Spectra.}

Numerical computations of theoretical radiative fluxes $F_{\nu}$
were carried out within the same classical 
approach as our model atmospheres using the WITA6 program \citep{Pavlenko2000}.
The shape of every line was determined using the Voigt function $H(a,v)$.
We account for natural broadening as well as van der Waals 
and Stark broadening of lines. Due to the low temperature and densities of 
C-giant model atmospheres Stark broadening may be ignored. Radiative 
broadening may play a major role in the 
outermost layers; in deep photospheric layers van der Waals broadening 
becomes important. 
Damping constants were taken from line databases such as VALD \citep{Kupka1999}
 or 
computed following \citet{Unsold1955}. 
The generation of our synthetic spectra are performed using the 
same opacity source lists and sets of abundances, as our model 
atmospheres. This approach allows us to directly see 
the impact of abundance changes on the temperature
structure of the model atmosphere as well as the synthetic spectra.
Computations of synthetic spectra   
were carried out with wavelength steps of 0.05 nm. 

\subsection{Fit to observed spectra}

There are several parameters that can be adjusted to fit synthetic spectra to observed stellar spectra. The most important of these include the logarithm of the carbon to oxygen ratio (\CO), effective temperature (\teff), surface gravity ($\log g$), metallicity ($\log(Z/Z_\odot)$) and turbulent velocity (\Vt). Fits obtained by varying these parameters are often not unique, for example \citet{Jorgensen2000} obtained two good fits for TX Psc, with different values of these parameters. In this work we attempt to gain good fits to the observed stellar spectra with \CO  and \Teff, but if necessary we change \Vt\  and the metallicity.

To obtain a fit we fix $\log g$ to 0.0, this value was estimated by \citet{Aoki1} from a measured radius of 200 R$_\odot$ for TX~Psc and an assumed mass of between 1 and 2 M$_\odot$, 
We also fix \Vt\  to 3.5 km s$^{-1}$ and metallicity to solar values, then calculate synthetic spectra on a \CO, \Teff\  grid. When a reasonable fit is found we adjust \CO\  and \Teff\  to obtain a better fit and if necessary change \Vt\  and metallicity to further improve the fit. We did not find it necessary to change $\log g$ to obtain good fits. Normalisation of the synthetic spectra and stellar spectra was carried out by dividing through by the Planck function of the relevant \Teff, then scaling the spectra so that the maximum flux is 1. The temperature of the black body spectrum used to normalise the stellar spectra is the \Teff\  of the best fit synthetic spectrum. The division of the observed spectra by the Planck function is adequate to normalise the spectrum to an pseudo-continuum.

\section{Results}
\label{sec:Disc}

The calculations reported here are the first to use the new HCN/HNC data of \cite{linepaper}. The fact that the data of \cite{linepaper} contains more lines implies a greater atmospheric opacity than accounted for in preceding HCN/HNC data \citep{Jorgensen1,Aoki1}. This increased opacity may have a significant effect on both the structure and spectra of our model atmospheres and synthetic spectra.

\subsection{Model Atmospheres}
\label{sec:models}

\begin{figure*}
%\epsscale{0.9}
%\plotone{PvsT_T.ps}
%\includegraphics{PvsT_T.eps}
\includegraphics{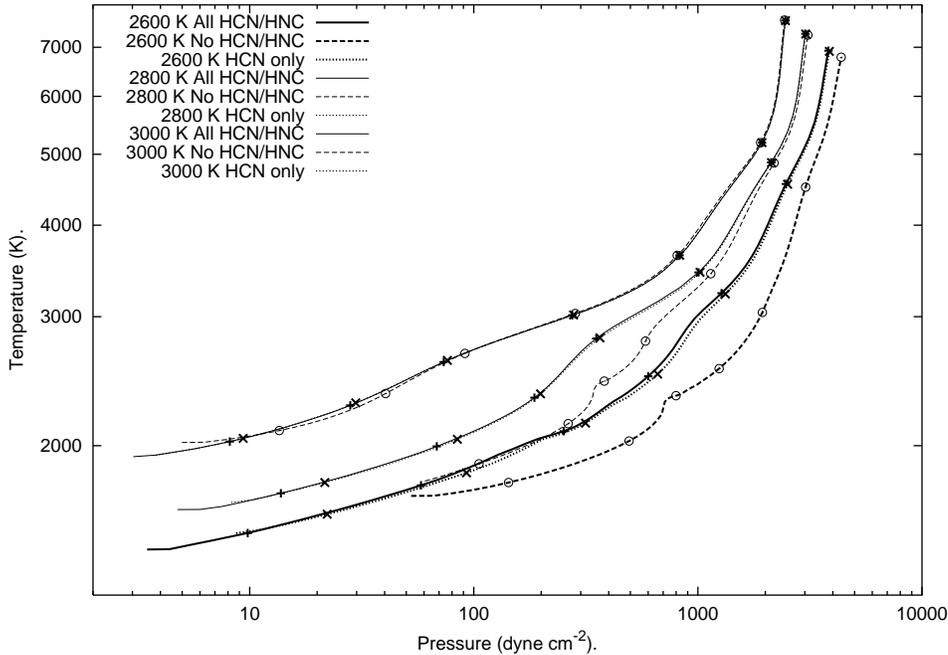}
\caption{Plot of pressure versus temperature on logarithmic scales for models with HCN/HNC lines (solid, $+$ symbols), with HCN lines only (dotted, $\times$ symbols) and without any HCN/HNC lines (dashed, $\circ$ symbols), at T=2600 (thick), 2800 (medium) and 3000 K (thin), with \CO\  of 0.01. Symbols are placed at Rosseland optical depths of $\log \tau_R$=-4, -3, -2, -1, 0, 1, 2, 3.}
\label{fig:PvsT_T}
\end{figure*}

Figure \ref{fig:PvsT_T} shows a plot of pressure versus temperature for 9 different model atmospheres with a \CO\  of 0.01, temperatures of 2600, 2800 and 3000 K, and with HCN/HNC opacity with only HCN opacity and without HCN/HNC opacity. Symbols are placed on the curves at value of $\log \tau_R$ of -4, -3, -2, -1, 0, 1 and 2. The difference between the model atmospheres with and without HCN/HNC opacity at 2800 and 2600 K is considerable. At a given optical depth the atmospheres with HCN/HNC opacity are cooler and at a much lower pressure, which is consistent with the trends originally reported by \citet{Eriksson}. 
However in the model atmospheres with an effective temperature of 3000 K, most of the HCN/HNC has been destroyed and the differences between spectra with and without HCN/HNC opacity is much reduced. None-the-less these differences are still significant, especially in the outer parts of the atmosphere. 
The model atmospheres with HCN/HNC and with only HCN, have discernible differences. At a given optical depth the models with HCN opacity only are hotter and at higher pressures than the models with HCN/HNC opacity. Much of this change in the optical depth pressure-temperature relationship is due to the increase in opacity and hence optical depth as a result of HNC absorption. However some of the differences are also attributable to changes in the pressure temperature curve of the atmospheres, brought about by increased absorption. The change in the pressure temperature curve is most pronounced in the 2600 K models.

%fig2 PvsT_T

%Interestingly at this effective temperature the model atmosphere including HCN/HNC opacity is slightly cooler and at higher pressure than the model without HCN/HNC opacity. This opposite to, but smaller than the effect seen in the cooler models .
%In these warmer atmospheres the temperature gradient is larger, consequently HCN only exist in a narrow region. This layer is to small to warm the entire atmosphere, but does screen the radiation field so that the temperature drops. Conversely in a cooler atmosphere the layers in which HCN exists are wider and the HCN number densities greater. Hence the effect is a warming of the entire atmosphere.

\begin{figure*}
%\epsscale{0.9}
%\plotone{PvsT_co.ps}
%\includegraphics{PvsT_co.eps}
\includegraphics{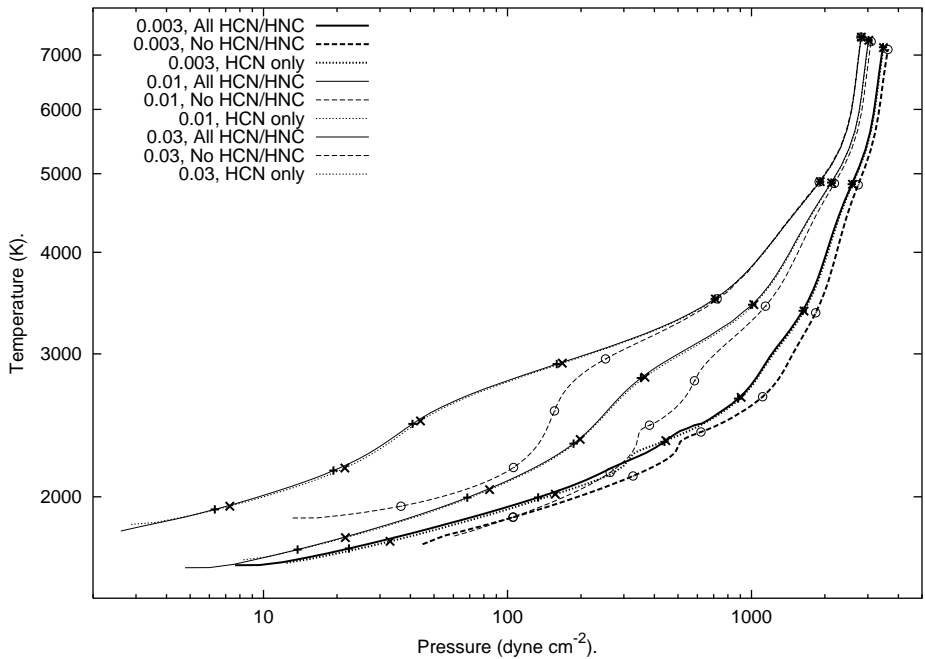}
\caption{Plot of pressure versus temperature on logarithmic scales for models with HCN/HNC lines (solid lines, $+$ symbols), with HCN lines only (dotted lines, $\times$ symbols) and without any HCN/HNC lines (dashed, $\circ$ symbols), at T=2800, with \CO\ of 0.003 (thick), 0.01 (medium) and 0.03 (thin). Symbols are placed at Rosseland optical depths of $\log \tau_R$=-4, -3, -2, -1, 0, 1, 2, 3.}
\label{fig:PvsT_co}
\end{figure*}

Figure \ref{fig:PvsT_co} shows a plot of pressure versus temperature for 9 model atmospheres with \CO\  of 0.003, 0.01 and 0.03 at a temperature of 2800 K, with and without HCN/HNC opacity and with HCN opacity only. As in figure \ref{fig:PvsT_T} the model atmospheres with HCN/HNC opacity are cooler and at much lower pressure than those without HCN/HNC opacity. It is also clear that the atmospheres with a higher C/O ratio have a hotter lower pressure atmosphere, this is consistent with the findings of \citet{Aoki1}. The reason is that more carbon rich molecules can form at higher C/O ratio resulting in a more opaque, so hotter and less dense, atmosphere. All these pressure-temperature curves are consistent with the model atmospheres reported by \citet{Aoki1} and \citet{Jorgensen2000}. Again there are differences between the models which account for HCN/HNC opacity and the models which account for only HCN opacity. 

% fig3 PvsT_co

\subsection{Spectra}

\begin{figure*}
%\epsscale{0.9}
%\plotone{synth_T.ps}
\includegraphics{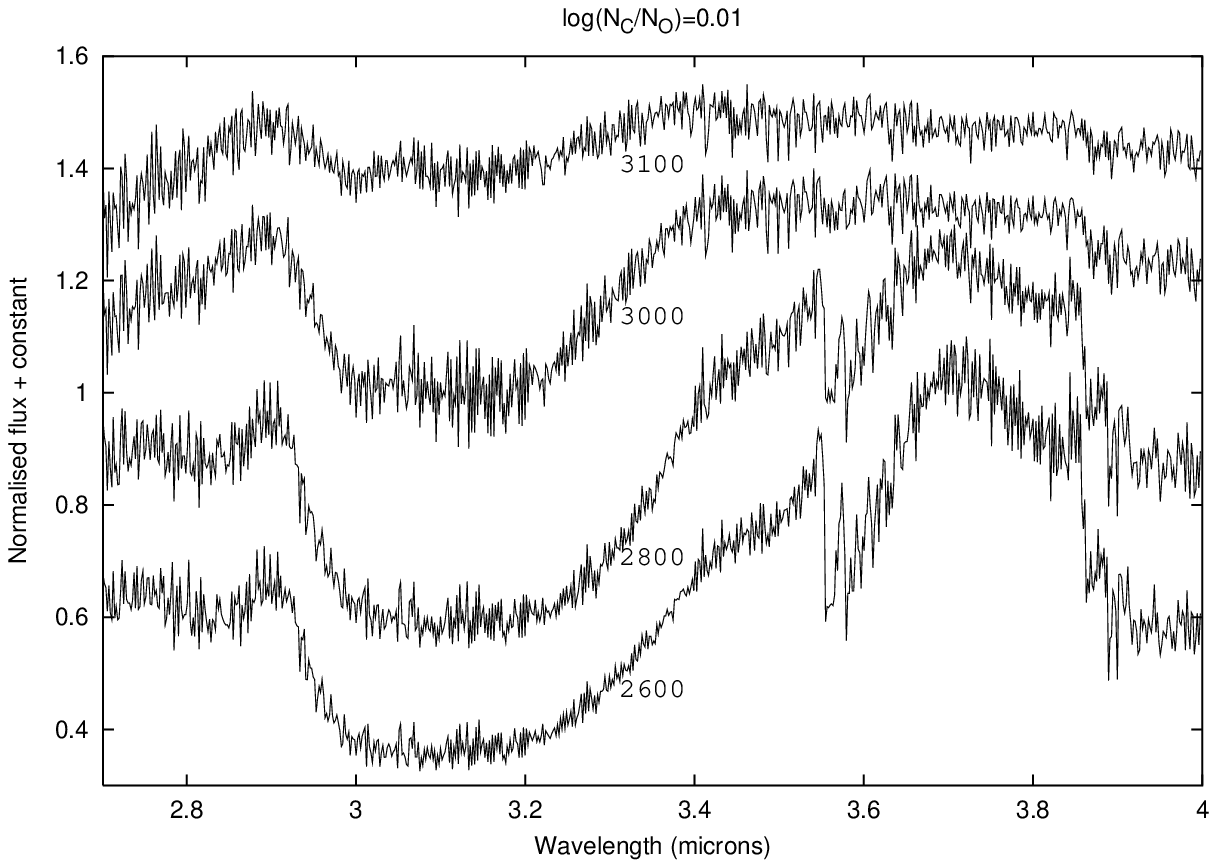}
\caption{Synthetic spectra with a \CO\  of 0.01 and effective temperatures of 2600, 2800, 3000 and 3100 K.}
\label{fig:synth_T}
\end{figure*}

Figure \ref{fig:synth_T} shows synthetic spectra calculated at effective temperatures of 2600, 2800, 3000, 3100 K. At \Teff of 2600 and 2800 K the atmospheric temperatures and densities are conducive to the formation of HCN/HNC, so the spectra between 2.7 and 4.0 \mim\  are dominated by HCN absorption.
Between 2.9 to 3.4 \mim\  is the characteristic strong absorption feature which is predominately caused by the HCN, H-C stretching mode. There are strong narrow absorption features around 3.6 \mim\  caused by HCN Q branches, see below. Immediately either side of the 3.6 \mim\  feature are broader weaker absorption features corresponding to P and R branches of the same bands. There are also strong, narrow absorption features longwards of 3.8 \mim, caused by a combination of CS and HNC bands, see below. 
Finally the broad absorption around 2.8 \mim\  is caused by the HNC, H-N stretching fundamental and its hot bands. 
At T$_{\rm eff}$ = 3000 K, all these absorption features are weaker, the narrow features at 3.6 \mim\  and longwards of 3.8 \mim\  have all but vanished. The 3 to 3.4 \mim\  feature has narrowed and weakened. The HNC absorption at 2.8 \mim\  is considerably weaker, and the spectrum in this region is becoming dominated by CO. At \teff=3200 K, almost no trace of HCN and HNC absorption is left, even at 3  \mim. Longwards of 2.9 \mim\  the spectrum is flat. These changes are a direct result of the temperature dependence of the HCN and HNC abundance.
As the effective temperature increases to 3000 K the HCN/HNC molecules begin dissociating to form CN, this process is complete by 3200 K at which point little sign of HCN is present.

% fig4, synth_T

\begin{figure*}
%\epsscale{0.9}
\includegraphics{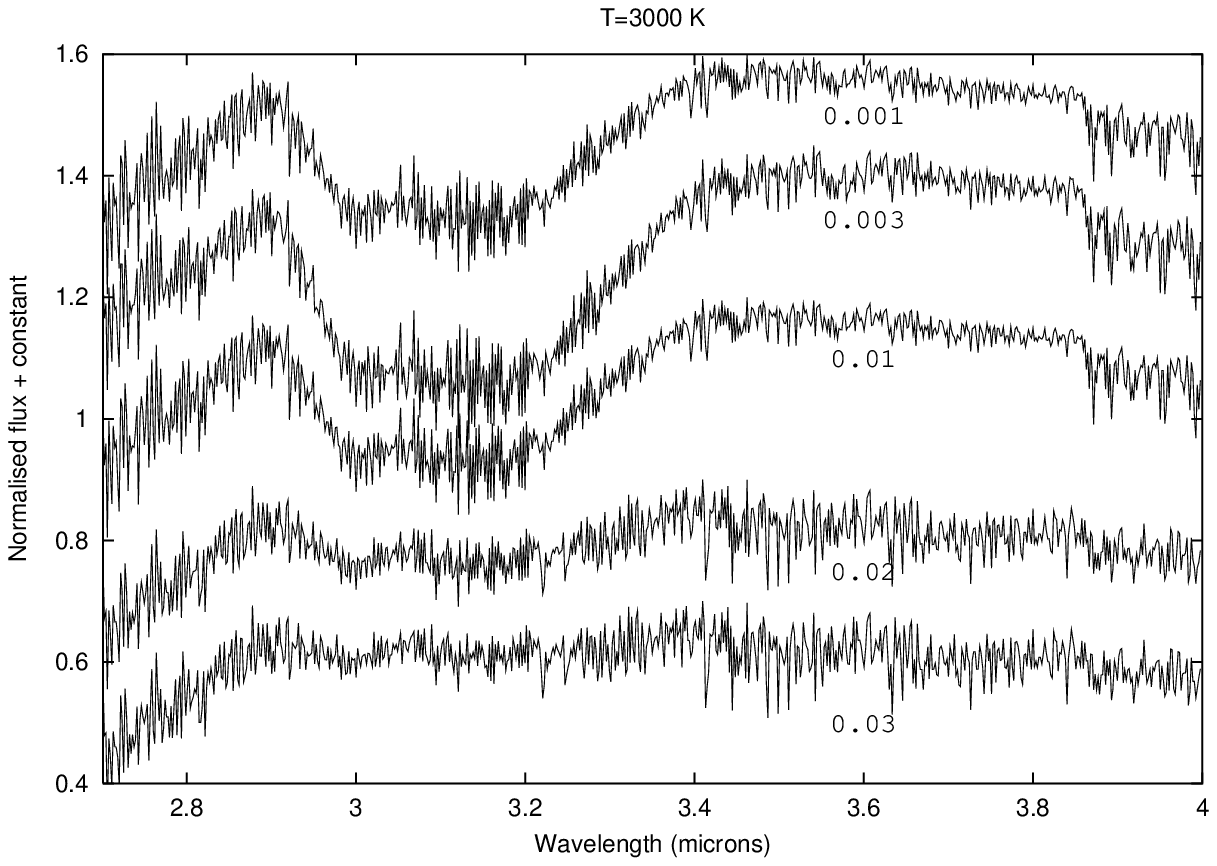}
%\includegraphics{synth_co.eps}
%\plotone{synth_co.eps}
\caption{Synthetic spectra with an effective temperature of 3000 K and \CO\  of 0.03, 0.02, 0.01, 0.003 and 0.001. A constant has been added to the flux of some of the spectra, to stagger them.}
\label{fig:synth_co}
\end{figure*}

Figure \ref{fig:synth_co} shows synthetic spectra calculated with an effective temperature of 3000 K and with \CO\  of 0.03, 0.02, 0.01, 0.003 and 0.001. At \teff=3000 K the HCN/HNC abundance is highly sensitive to both the C/O ratio and temperature. This is seen in both figures \ref{fig:synth_T} and \ref{fig:synth_co}. The 3-3.4 \mim\  feature is strong for \CO\  between 0.001 and 0.01, but is weak at \CO$=$ 0.02 and almost no existent at \CO$=$0.03. Here higher C/O ratios give rise to a greater abundance of carbon rich molecules which in turn makes the atmosphere more opaque and so hotter and less dense, see figure \ref{fig:PvsT_co}. Such conditions do not favour the formation of HCN/HNC. So a high C/O ratio can result in the destruction of HCN/HNC. These sensitivities to the C/O ratio at \teff$\sim$3000 K, make it difficult to fit spectra to the higher temperature C-stars, such as TX Psc, see section \ref{sec:TXPsc_fit}.

% fig5, synth_co

As discussed in section \ref{sec:Star} we have obtained suitable spectra of the carbon rich AGB stars WZ Cas and TX Psc, from the ISO archives. The inclusion of the new extensive HCN/HNC linelist \citep{linepaper} within our synthetic spectra allows the identification of HCN and HNC features, within these stellar spectra.
The new HCN/HNC are more complete and in the case of ab initio data more accurate, than the preceding HCN/HNC data and so allow the determination of new effective temperatures and C/O ratio for these stars.

\subsubsection{WZ Cas.}
\label{sec:WZCas_fit}

\begin{figure*}
%\epsscale{0.9}
\includegraphics{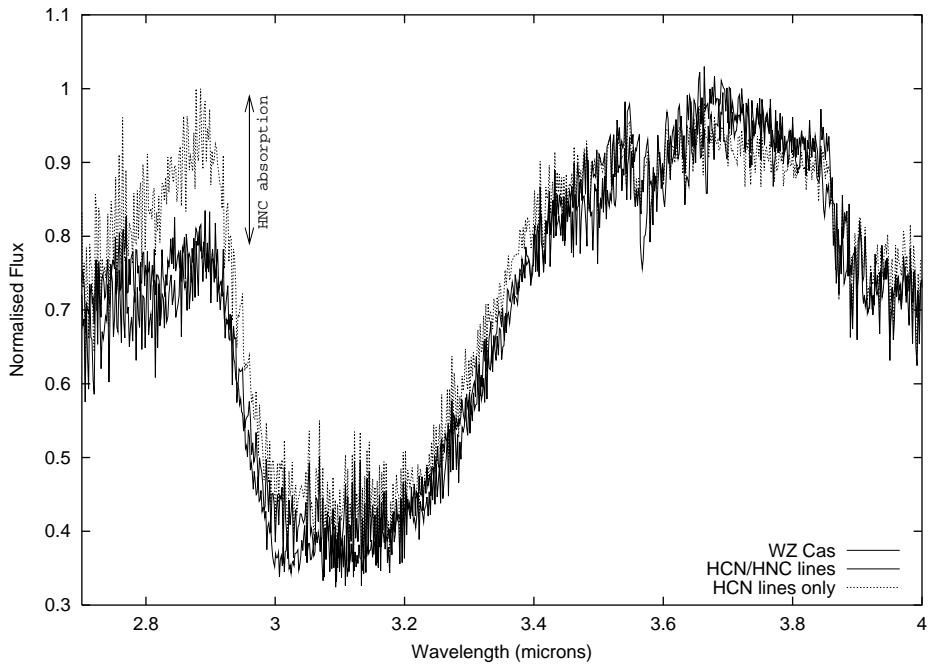}
%\includegraphics{bestfit_WZCAS.eps}
%\plotone{bestfit_WZCAS.ps}
\caption{Synthetic spectra of carbon rich atmospheres at T$_{\rm eff}=$2800 K, \CO\  of 0.003, with a V$_t$=3.5 km s$^{-1}$ in the region 2.7-4.0 $\mu$m. The spectra were calculated with the full HCN/HNC linelist of \citet{linepaper} and with only HCN lines selected. For comparison the observed spectrum of WZ Cas is shown. The resolution of these spectra is 1.6 nm.}
\label{fig:bestfit_WZCAS}
\end{figure*}

Figure \ref{fig:bestfit_WZCAS}, shows the spectrum of WZ~Cas together with the best fit synthetic spectrum calculated with HCN/HNC opacity and a synthetic spectrum calculated using only HCN opacity. Both of the synthetic spectra were computed for a model atmosphere of \Teff =2800 K, a turbulent velocity of 3.5 km s$^{-1}$ and a value of \CO\  of 0.003. To permit easy identification of the broad absorption features within the spectra, both synthetic and observed spectra are shown at a resolution of 1.5 nm. Recent measurements and fits for the effective temperature and \CO\  are shown in table \ref{tab:Teff_WZCas}. There are large differences between the values of \Teff. The estimates of \Teff\  from                      \citet*{Dyck} is 350 K hotter than the estimate of \Teff\  from our models. However if we use this \Teff\  in our models, HCN does not form in large enough quantities to form strong absorption features, see figure \ref{fig:synth_T}. There are large errors of 200 K, on the value of \Teff\  quoted by \citet{Dyck}. Our estimate of the \Teff\  and \CO\  for WZ Cas is closer to that of \citet{Lambert} and that of \citet{Aoki1} which is an recalibration of the infrared flux method (IRFM) \citep*{Blackwell} determination by \citet{Ohnaka96}.

% fig6 bestfit_WZCAS AND table1

\begin{table*}
 \centering
 \begin{minipage}{140mm}
\caption{Effective temperatures, C/O ratios and surface gravities for WZ Cas.}
\begin{tabular}{lclrl}
\hline
Work &     		\teff\	& \CO   & log(g) & Notes \\ \hline

\citet{Lambert}         & 2850          & 0.0043 &     & \Teff\  from colour index's, \CO\  from models. \\

\citet{Dyck}            & 3140$\pm$193  &       &      & interferometry \\

\citet{Aoki1}           & 2900	& 0.0043 & 0.0  & Recalibrated IRFM determination of \citet{Ohnaka96} \\
 This work 	& 2800	& 0.003 & 0.0  & model \\

\end{tabular}
\label{tab:Teff_WZCas}
\end{minipage}
\end{table*}

In general the major features of the synthetic spectrum with both HCN and HNC opacity, fit well the observed spectrum of WZ Cas. 
The absorption feature above 3.8 microns, is attributable to a combination of the Q branch absorption of the (0,1,1,0)$\rightarrow$(1,0,0,0) band together with its hot bands and the $\Delta v=2$ bands of CS. 
The general shape of this feature is well reproduced in both synthetic spectra. However the positions of the individual Q branches in the synthetic spectra are slightly displaced from the same Q branches in the observed spectrum. This is a result of the errors on the line frequencies in the ab initio data of \citet{linepaper}. There is slightly weaker absorption in this region in the synthetic spectrum with only HCN absorption than there is in the spectrum which uses the full HCN/HNC opacity. This is most likely a result of the effect of absorption by the HNC (0,0,0,0)$\rightarrow$(0,1,1,1) band, centred on 4.03 \mim , and its hot bands. From figure \ref{fig:HCN_abs} it is expected that HNC will contribute to the overall HCN/HNC opacity in this region. Some of the changes in the synthetic spectra may also have been brought about by changes in the optical depth, pressure and temperature of the model atmosphere.

The absorption in the 3.6 \mim\  region is attributable to the Q branches of the HCN (0,0,0,0)$\rightarrow$(0,1,1,1) band.
Again the general shape of this feature in both synthetic spectra is well reproduced, but the Q branches in the synthetic spectra are displaced from the observed spectra, a result of the error on the ab initio HCN/HNC data of \citet{linepaper}. 
Shortwards of 3.55 \mim\  the spectrum which contains both HCN and HNC absorption agrees well with the observed spectra. The spectrum which does not contain HNC data appears to be missing absorption, especially in the 2.9 \mim\  region. The broad absorption feature from 2.95 to 3.4 \mim\  which is widely attributed to the HCN H-C stretching fundamental and hot bands, centred on 3.02 \mim, and its hot bands. This region has been well documented in the past (for example \citet{Aoki1})  . There should also be a little absorption from the C$_2$H$_2$ symmetric and anti-symmetric H-C stretch modes. However the difference between the intensity of this feature in the synthetic spectra with and without HNC, suggest that much absorption in this region is a result of HNC. 
The strongest HNC mode in this region is the (0,1,1,0)$\rightarrow$(1,0,0,0) band, centred on 3.125 \mim, and its hot bands. The effect of this band on the combined HCN/HNC opacity can be seen in figure \ref{fig:HCN_abs}.

The biggest difference between the synthetic spectra with HCN/HNC lines and with HCN lines only, lies in the 2.85 \mim\  region. The dominant absorption in this region of the synthetic spectrum with only HCN opacity, is caused by CO absorption from the $\Delta v=2$ band. However, in the synthetic spectrum which includes the full HCN/HNC opacity there is additional absorption. This is the HNC (0,0,0,0)$\rightarrow$(1,0,0,0) fundamental and hot bands, centred on 2.73 \mim.
From the ab initio absorption spectrum of HCN/HNC (figure  \ref{fig:HCN_abs}) it would be expected that if HCN and HNC are in thermodynamic equilibrium there should be significant HNC absorption in this region. In the 2.85 \mim\  region the difference between the observed and synthetic spectra are also at their greatest. The synthetic spectrum which includes the full HCN/HNC linelist fits both the shape and flux of the spectrum of WZ Cas considerably better than does the synthetic spectrum which contains only HCN absorption. Thus it is here that we identify the effect of HNC absorption in the spectrum of a carbon star for the first time. 

The synthetic spectra of \citet{Aoki1} were calculated using available laboratory HCN data. \citet{Aoki1} were not able to fully account for the HCN absorption between 3.8 to 4.2 \mim, using only this laboratory data.
To gain a closer agreement with observed spectra \citet{Aoki1} had to empirically estimate opacity from $\Delta v_1=2$, $\Delta v_2=-1$  hot bands with $v_2 > 5$. Our synthetic spectra do not suffer from this data deficit and are on the whole are in good agreement in this region. The minimum frequency of the synthetic spectra of \citet{Aoki1} is 2.9 \mim, so we are unable to comment upon the spectra at 2.8 \mim\  where HNC is a strong absorber.

\begin{figure*}
%\epsscale{0.8}
\includegraphics{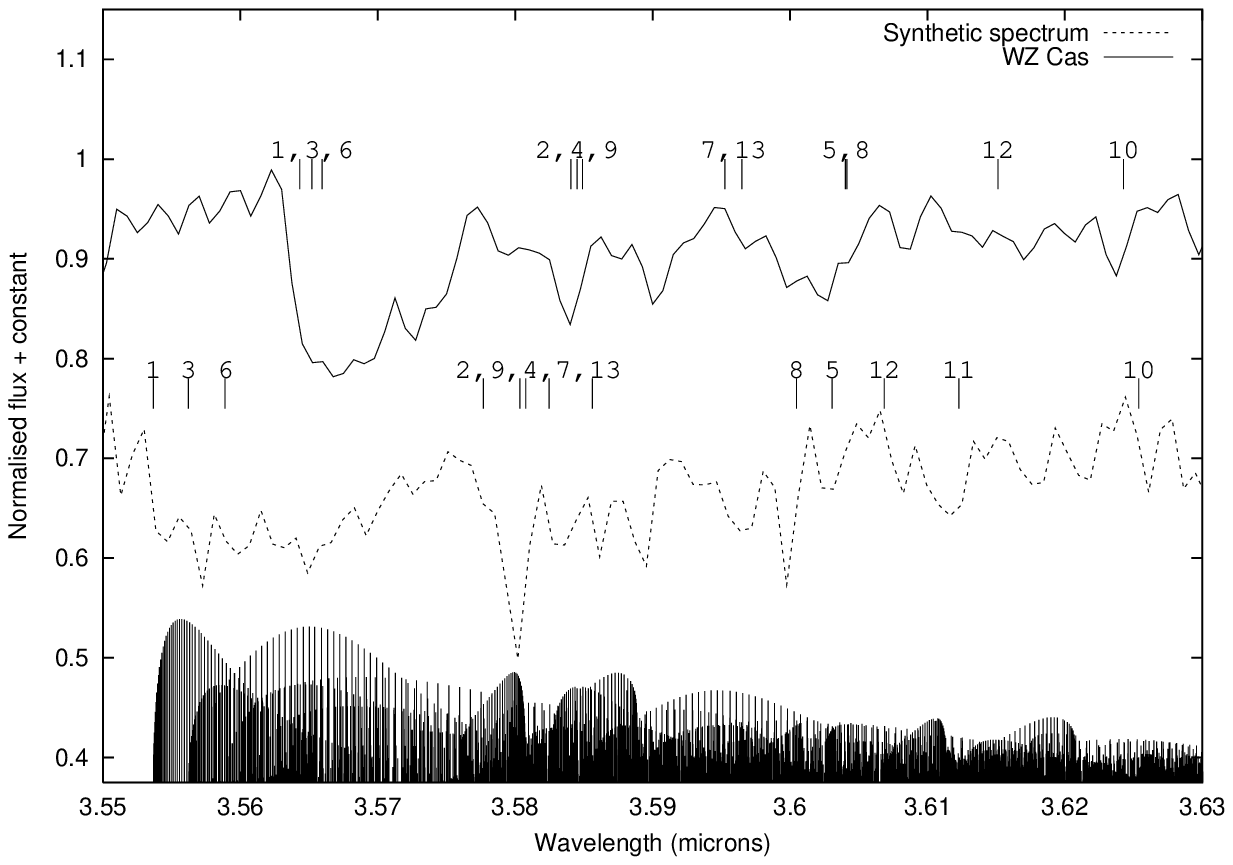}
%\includegraphics{assign_3.55-3.61.eps}
%\plotone{assign_3.55-3.61.ps}
\caption{The observed spectrum of WZ Cas over the region of 3.55-3.61 $\mu$m, with a synthetic spectrum at T=2800 K, \CO\  of 0.003 and $V_T=3.5$ km s$^{-1}$. A constant is subtracted from the flux of the synthetic spectrum to stagger it from the observed spectrum. The Q branch lines in this spectral region in the ab initio data of \citet{linepaper} are plotted at the bottom of the figure with line length in proportion to intensity. The experimental band centres of the (0,0,0,0)$\rightarrow$(0,1,1,1) band and its hot bands are indicated by the upper series of marks and digits. Likewise the lower series of marks and digits identify the band centres of the ab initio data of \citet{linepaper}. The data for these bands, together with its index are tabulated in table \ref{tab:0111-0000}. }
\label{fig:assign_3.55-3.61}
\end{figure*}

\begin{figure*}
%\epsscale{0.8}
%\includegraphics{assign_3.8-4.0.eps}
\includegraphics{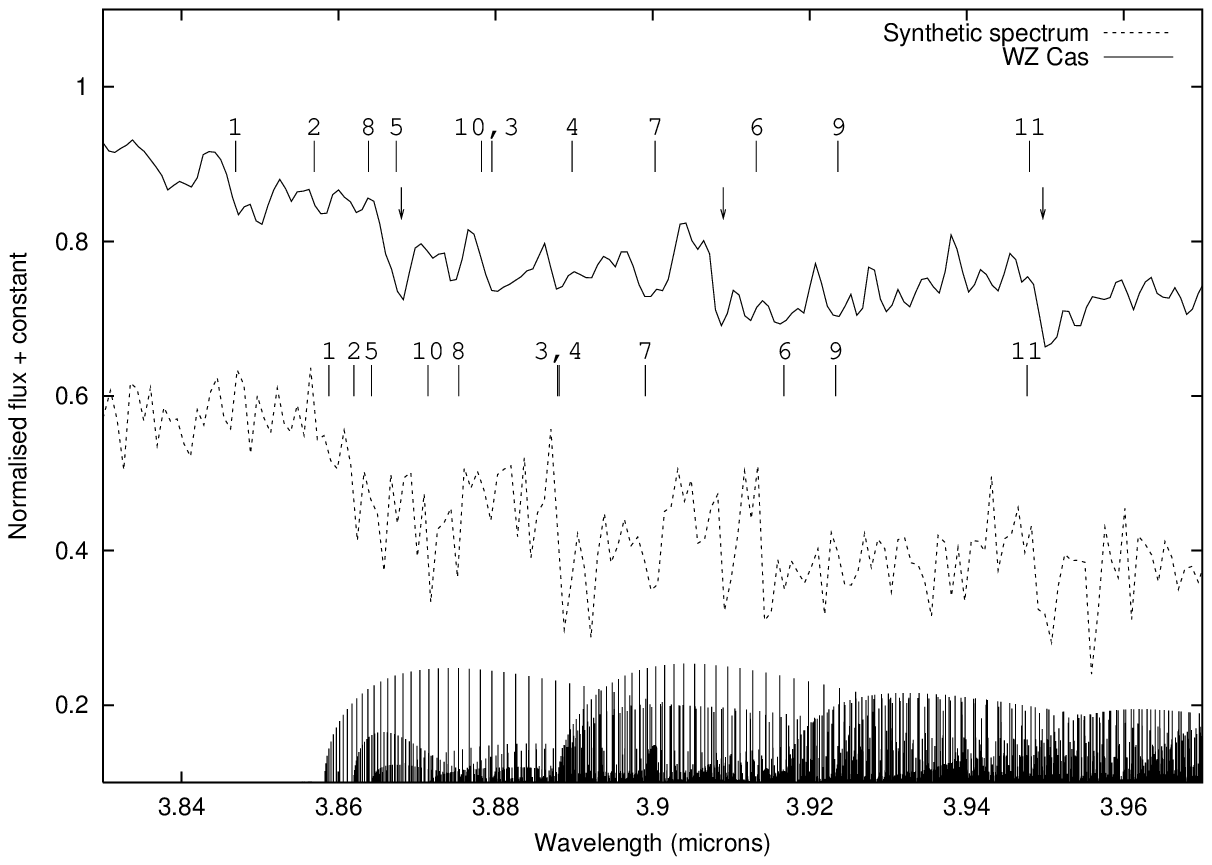}
%\plotone{assign_3.8-4.0.ps}
\caption{The observed spectrum of WZ Cas over the region of 3.8-4.0 $\mu$m, with a synthetic spectrum at T=2800 K, \CO\  of 0.003 and $V_T=3.5$ km s$^{-1}$. A constant is subtracted from the flux of the synthetic spectrum to stagger it from the observed spectrum. The Q branch lines in this spectral region in the ab initio data of \citet{linepaper} are plotted at the bottom of the figure with line length in proportion to line intensity. The experimental band centres of the (0,1,1,0)$\rightarrow$(1,0,0,0) band and its hot bands are indicated by the upper series of marks and digits. Likewise the lower series of marks and digits identify the band centres of the ab initio data of \citet{linepaper}. The downward pointing arrows indicate the position of CS band heads. The data for the HCN bands with their index is tabulated in table \ref{tab:1000-0110}. }
\label{fig:assign_3.8-4.0}
\end{figure*}

For the spectral ranges 3.55 to 3.63 \mim\  and 3.83 to 3.97 \mim\  we plot the 
observed spectrum of WZ Cas together with the best fit synthetic spectrum in figures \ref{fig:assign_3.55-3.61} and \ref{fig:assign_3.8-4.0}, respectively.
To help the reader identify the bands of HCN responsible for the absorption features in the synthetic and hence observed spectra we have plotted the ab initio HCN Q branch lines at the bottom of these figures. The band centres of the ab initio data of \citet{linepaper} are identified by the lower marks and digits and the experimentally determined band centres by the upper marks and digits. The indexes of the band centres are given in tables \ref{tab:0111-0000} and table \ref{tab:1000-0110}. To show the relatively broad absorption features the resolution of both the observed and synthetic spectra has been set to 0.75 nm.
The typical spacing between Q branch lines in triatomics is considerably smaller (about 0.1 nm) than the spacing between P and R branch lines (about 4 nm). Consequently, the intensity of the P and R branches is distributed over a large wavelength range. Hence the P and R branches are broader and weaker than the Q branches which are characterised by narrow and strong absorption. 
The effects of Q branches in the spectra are thus often more apparent than the effects of P and R branches. This is seen in both figures \ref{fig:assign_3.55-3.61} and \ref{fig:assign_3.8-4.0} . 
The absorption features correspond to the places in which the Q branch lines are densest. Usually this corresponds to the position of the band centres. In some cases the line positions of a Q branch lines will increase in wavelength with J and at some point reverse and decrease in wavelength with J, in a similar fashion to a band head. For example the feature at about 3.58 \mim\  in the synthetic spectrum is attributable to a Q branch of the (0,3,1f,1)$\rightarrow$(0,4,2e,1) band which displays such behaviour. 

% fig7 assign_3.55-3.61 and TABLES 3, 4

\begin{table*}
 \centering
 \begin{minipage}{140mm}
\caption{Laboratory and ab initio  band centres for bands with $\Delta v_2=1$,$\Delta v_3=1$.}
\begin{tabular}{ccccccc}
\hline
index \footnote{This is the index of the band used in figure \ref{fig:assign_3.55-3.61}} &
$(v_1^\prime,v_2^\prime,l^\prime,v_3^\prime)$ & 
$(v_1^{\prime\prime},v_2^{\prime\prime},l^{\prime\prime},v_3^{\prime\prime})$ & 
$E_c^\prime$ \cm & $E_c^{\prime\prime}$ \cm  & Theory \footnote{Calculated from the ab initio data of \citet{linepaper}.}  $\lambda_c$ \mim & Exp \footnote{From the laboratory measurements of \citet{Maki96}.}
  $\lambda_c$ \mim \\ \hline

 1 & (0,1,1,1)  & (0,0,0,0) & 2814.00 &    0.0  & 3.5537 & 3.5643 \\
 2 & (0,2,0,1)  & (0,1,1,0) & 3510.99 &  715.89 & 3.5777 & 3.5841 \\
 3 & (0,2,2,1)  & (0,1,1,0) & 3527.86 &  715.89 & 3.5562 & 3.5652 \\
 4 & (0,3,1,1)  & (0,2,0,0) & 4207.62 & 1414.92 & 3.5808 & 3.5845 \\
 5 & (0,3,1,1)  & (0,2,2,0) & 4207.62 & 1432.20 & 3.6031 & 3.6040 \\
 6 & (0,3,3,1)  & (0,2,2,0) & 4242.07 & 1432.20 & 3.5589 & 3.5659 \\
 7 & (0,1,1,2)  & (0,0,0,1) & 4891.95 & 2100.58 & 3.5825 & 3.5953 \\
 8 & (0,4,0,1)  & (0,3,1,0) & 4891.76 & 2114.34 & 3.6005 & 3.6041 \\
 9 & (0,4,2,1)  & (0,3,1,0) & 4907.37 & 2114.34 & 3.5803 & 3.5849 \\
10 & (0,4,2,1)  & (0,3,3,0) & 4907.37 & 2149.03 & 3.6254 & 3.6243 \\
11 & (0,4,4,1)  & (0,3,3,0) & 4957.36 & 2149.03 & 3.6123 & -- \\
12 & (0,2,0,2)  & (0,1,1,1) & 5586.50 & 2814.00 & 3.6069 & 3.6151 \\
13 & (0,2,2,2)  & (0,1,1,1) & 5602.92 & 2814.00 & 3.5856 & 3.5965 \\

\end{tabular}

\label{tab:0111-0000}
\end{minipage}
\end{table*}

\begin{table*}
 \centering
 \begin{minipage}{140mm}
\caption{Laboratory and ab initio  band centres for bands with $\Delta v_1=1$,$\Delta v_2=-1$.}
\begin{tabular}{ccccccc}
\hline
index \footnote{This is the index of the band used in figure \ref{fig:assign_3.8-4.0}} &
$(v_1^\prime,v_2^\prime,l^\prime,v_3^\prime)$ & 
$(v_1^{\prime\prime},v_2^{\prime\prime},l^{\prime\prime},v_3^{\prime\prime})$ & 
$E_c^\prime$ \cm & $E_c^{\prime\prime}$ \cm  & Theory \footnote{Calculated from the ab initio data of \citet{linepaper}.}\  $\lambda_c$ \mim & Exp\footnote{From the laboratory measurements of \citet{Maki96}.}
\  $\lambda_c$ \mim \\ \hline

 1 & (1,0,0,0)  & (0,1,1,0) & 3307.70 &    0.0  & 3.8588 & 3.8469 \\
 2 & (1,1,1,0)  & (0,2,0,0) & 4004.28 & 1414.92 & 3.8620 & 3.8569 \\
 3 & (1,1,1,0)  & (0,2,2,0) & 4004.28 & 1432.20 & 3.8879 & 3.8795 \\
 4 & (1,2,0,0)  & (0,3,1,0) & 4686.28 & 2114.34 & 3.8881 & 3.8898 \\
 5 & (1,2,2,0)  & (0,3,1,0) & 4702.19 & 2114.34 & 3.8642 & 3.8673 \\
 6 & (1,2,2,0)  & (0,3,3,0) & 4702.19 & 2149.03 & 3.9167 & 3.9132 \\
 7 & (1,3,1,0)  & (0,4,0,0) & 5366.17 & 2801.46 & 3.8991 & 3.9003 \\
 8 & (1,0,0,1)  & (0,1,1,1) & 5394.43 & 2814.00 & 3.8753 & 3.8638 \\
 9 & (1,3,1,0)  & (0,4,2,0) & 5366.17 & 2817.27 & 3.9233 & 3.9236 \\
10 & (1,3,3,0)  & (0,4,2,0) & 5400.32 & 2817.27 & 3.8714 & 3.8782 \\
11 & (1,3,3,0)  & (0,4,4,0) & 5400.32 & 2867.21 & 3.9477 & 3.9480 \\

\end{tabular}

\label{tab:1000-0110}
\end{minipage}
\end{table*}

This comparison of the Q branch line positions and intensities in figures  \ref{fig:assign_3.55-3.61} and \ref{fig:assign_3.8-4.0} with the synthetic spectra allows us to attribute some of the absorption features in the observed spectra to Q branches. In figure \ref{fig:assign_3.55-3.61}
the deep absorption at 3.567 \mim\  in the spectrum of WZ Cas is attributable to the Q branches of the (0,0,0,0)$\rightarrow$(0,1,1,1), (0,1,1,0)$\rightarrow$(0,2,2,1), (0,2,2,0)$\rightarrow$(0,3,3,1) bands and higher hotbands. 
The feature at 3.584 \mim\  is attributable in the most part to the Q branches of the (0,2,0,0)$\rightarrow$(0,3,1,1) and the (0,3,1,0)$\rightarrow$(0,4,2,0) bands. The line spacing of the (0,1,1,0)$\rightarrow$(0,2,0,1) is large and 
consequently the intensity of the Q branch is spread over a larger wavelength range so is less detectable. 
Beyond 3.585 \mim\ the increasing density of bands coupled with the errors on the ab initio data make it difficult to attribute absorption features to specific bands.
The synthetic spectra of \citet{Aoki1} reproduce the 3.56-3.63 \mim\  band, but because of the limitations of their linelist they were unable to fully explain the various features in this region.
The match of the synthetic spectra to the observed spectrum would be much improved if better frequency data was used.
It is possible to substitute experimentally determined HCN/HNC energy levels and hence line frequencies into the \citet{linepaper} linelist. This would improve the line frequencies for all the lines corresponding to transitions between experimentally determined energy levels. However, this would require the manual assignment of the ``approximate'' quantum numbers $v_1$, $v_2$, $v_3$ and $l$ to many of the 125~000 energy levels included in the \citet{linepaper} linelist. \citet{Bob} have already assigned about a quarter of the levels we needed to be able to insert experimentally determined energy levels.

The spectral region covered by figure \ref{fig:assign_3.8-4.0} is slightly more complex, because there is significant absorption from both HCN and CS. The band heads of the CS 0$\rightarrow$1, 1$\rightarrow$2 and 2$\rightarrow$3 lie at 3.868 \mim, 3.909 \mim\  and 3.9497 \mim, respectively \citep{Chandra}. Three absorption features in the observed and synthetic spectrum of WZ~Cas caused by these band heads are clearly visible and have previously been commented upon by \citet{Aoki1} in WZ~Cas and TX~Psc and also by \cite{Jorgensen2000} in TX~Psc.
As far as HCN absorption is concerned the 3.85 \mim\  feature in the spectrum of WZ Cas is attributable to the Q branch of the (0,1,1,0)$\rightarrow$(1,0,0,0) band. The absorption at about 3.88 microns can be confidently associated with the Q branches of the (0,2,2,0)$\rightarrow$(1,1,1,0) and (0,4,2,0)$\rightarrow$(1,3,3,0) bands. There is also a strong case for the 3.9 \mim\  absorption being caused by the unusually high density of Q branch lines in the (0,4,0,0)$\rightarrow$(1,3,1,0) band. It is also likely that the (0,2,0,0)$\rightarrow$(1,1,1,0) band is responsible for the weak absorption at 3.857 \mim\ in the spectrum of WZ Cas.
The increasing density and decreasing strength of both HCN and CS lines coupled with the errors of the ab initio HCN data hinder further identifications longwards of 3.905 \mim.

In this work we have assumed that the spectrum of H$^{13}$CN is identical to that of H$^{12}$CN. In reality this is not the case, the vibrational band centres and rotational constants are slightly different. \citet{Maki00} have measured band centres and rotational constants for H$^{13}$CN in the laboratory. These measurements found the band centres of the H$^{13}$CN fundamentals to be lower in wavenumber than those of H$^{12}$CN. The band centre of the fundamental of the C-N stretch mode differers by $\sim$ 1.6\% between the two isotopomers, the bending mode by $\sim$0.8\% and the H-C stretch mode by $\sim$0.5\%. If the abundance ratio of $^{13}$C to $^{12}$C is high in a star, then one would find that the absoption bands of HCN would have slightly different shapes. Furthermore there would be additional Q branches of H$^{13}$CN in the spectra, which would be slightly displace from those of H$^{12}$CN. When corrected for the change in center of mass the H$^{13}$CN electric dipole moment is identical to that of H$^{12}$CN. Consiquently the transition dipoles of the H$^{13}$CN should be close to those of H$^{12}$CN. To gain a good approxomation to the absorption of H$^{13}$CN the experimental energy levels of H$^{13}$CN could be subsituted for the theoretical levels of H$^{12}$CN in the \citet{linepaper} linelist. This, however, must also wait until sufficent energy levels of the \citet{linepaper} linelist are assigned.

% fig8 assign_3.8-4.0

\subsubsection{TX Psc.}
\label{sec:TXPsc_fit}

\begin{figure*}
%\epsscale{0.9}
\includegraphics{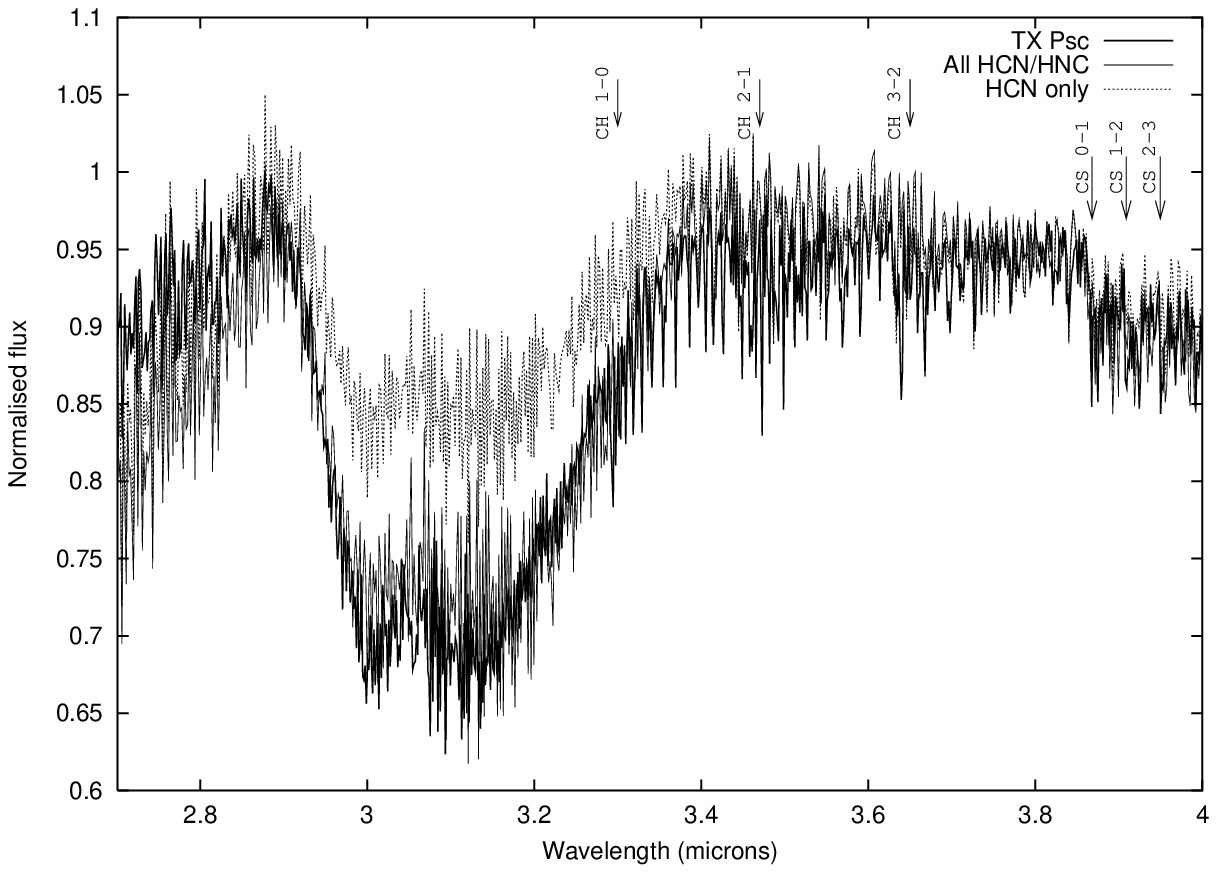}
%\includegraphics{bestfit_TXPSC.eps}
%\plotone{bestfit_TXPSC.ps}
\caption{Synthetic spectra of carbon rich atmospheres at T$_{\rm eff}=$3050 K with a V$_t$=3.5 km s$^{-1}$ and \CO=0.015 in the region 2.7-4.0 $\mu$m. The spectra were calculated with the full HCN/HNC linelist of \citet{linepaper} and with just HCN lines selected from the \citet{linepaper} linelist, see text. For comparison the observed spectrum of TX Psc is shown. We have also indicated with arrows the positions of CS and CH band heads. The resolution of these spectra is 1.6 nm.}
\label{fig:bestfit_TXPSC}
\end{figure*}

Figure \ref{fig:bestfit_TXPSC} shows the ISO SWS spectrum of TX Psc, taken by \citet{Aoki1}, with our best fit synthetic spectra. One of the synthetic spectra is calculated with the full HNC/HCN opacity and the other is calculated with only HCN opacity. Both of the synthetic spectra are calculated with \teff\ $=3050$ K, $\log({\rm C/O})=0.03$, $V_t=3.5$ km s$^{-1}$ and $\log(Z/Z_\odot)$=-0.3. 

% fig9  bestfit_TXPSC

The \Teff\  and C/O ratio estimations for TX Psc of  \citet{Aoki1}, \citet{Jorgensen2000} and those of this work are summarised in table \ref{tab:Teff_TXPsc}, together with interferometric estimates of \Teff. All the values of \Teff\  deduced from the fitting of synthetic spectra agree to within 100 K, but the values obtained by interferometric means are upto 200 K cooler. 
\citet{Jorgensen2000} reported that their estimate of the \teff\  of TX Psc changed from 3100 K to 3000 K between the two separate ISO observations one taken by themselves and one taken by \citet{Aoki1}. However such temperature variations are difficult to confirm as the error in \Teff\  of our fit is around 100 K.
The value of \CO=0.015 from our work is closer to the lower \CO=0.010 value of the two good fit of \citet{Jorgensen2000}, than the higher values of 0.041-0.068 \citep{Aoki1,Jorgensen2000,Ohnaka2000}.
In our synthetic spectra the HCN absorption features are very weak for values of \CO\  greater than 0.02, see figure \ref{fig:synth_co}. So we believe the lower \CO\  values for TX Psc are closer to the real values.

% table 4

\begin{table*}
 \centering
 \begin{minipage}{140mm}
\caption{Effective temperatures, \CO\  and surface gravities for TX Psc.}
\begin{tabular}{lclrl}
\hline
Work &     		\teff\	& \CO   & log(g) & Notes \\ \hline

\citet{Lambert}         &    3030      & 0.012 &     & \Teff\  from colour index, \CO\  from models. \\

\citet{Quir}            & 2805$\pm$126 &       &      & interferometry \\
\citet{Dyck}            & 2921$\pm$60  &       &      & interferometry \\
\citet{Aoki1} 		& 3100	& 0.041 & 0.0  & Recalibrated IRFM determination of \citet{Ohnaka96} \\
\citet{Jorgensen2000}	& 3100	& 0.01	& -0.5 & model \\
\citet{Jorgensen2000}	& 3100	& 0.041	& 0.0  & model \\
\citet{Jorgensen2000}	& 3000	& 0.01 & -0.5 & model \\
\citet{Jorgensen2000}	& 3000	& 0.041 & 0.0  & model \\
\citet{Ohnaka2000}      & 3000  & 0.03  &  0.0  & model \\
\citet{Ohnaka2000}      & 3100  & 0.068 &  0.0  & model \\  
\citet{Ohnaka2000}      & 3000  & 0.041 & -0.5  & model \\

 This work		& 3050	& 0.015 & 0.0  & model \\

\end{tabular}

\label{tab:Teff_TXPsc}
\end{minipage}
\end{table*}

There are dramatic differences between the two synthetic spectra shown in figure \ref{fig:bestfit_TXPSC}, the broad 3.1 \mim\  absorption feature in the synthetic spectrum with HCN and HNC opacity is much weaker  in the synthetic spectrum calculated with only HCN opacity. At \teff\ $=3000$ K the model atmospheres are very sensitive, slight changes in temperature and \CO\  can have a dramatic effect on the synthetic spectra, for example see figure \ref{fig:synth_T} and \ref{fig:synth_co}. As discussed in section \ref{sec:models}, a model without HNC opacity has a higher temperature and pressure at a given optical depth than a model with the full HCN/HNC opacity.
Hence a reduction in opacity caused by neglecting HNC opacity can affect the structure of the atmosphere to the extent at which HCN absorption is dramatically reduced. As in the case of WZ~Cas both HCN and HNC are required to explain the observed spectra. Marked on figure \ref{fig:bestfit_TXPSC} are the  0$\rightarrow$1, 1$\rightarrow$2 and 2$\rightarrow$3 CS and CH band heads, which were identified previously by \citet{Aoki1}.

Although the synthetic spectra calculated with HCN/HNC opacity fits the observed spectrum well between 2.9 to 4.0 \mim, there is a slight under-prediction of the strength of the absorption between 3.0 and 3.2 \mim. Between 2.7 and 2.9 \mim\  the flux from the synthetic spectrum dips to below the spectrum of TX Psc. The absorption across this region is effected by the tail of the CO $\Delta v=2$ bands. The effect of this CO band is evident in the synthetic spectra shown in figure \ref{fig:synth_co}. This would imply that there is too much CO within our model atmosphere. We have looked at other potential ways of reducing the abundance of CO, including changing the surface gravity and increasing and decreasing the value of \NO\  above and below the solar value. It is clear that both these parameters do affect the HCN absorption features in the final synthetic spectrum, but we have not found a better fit. The elemental abundances within carbon stars are not well known, so $\log(N_{\rm C}/N_{\rm H})$, $\log(N_{\rm N}/N_{\rm H})$, $\log(N_{\rm O}/N_{\rm H})$ or indeed $\log(N_{\rm He}/N_{\rm H})$  are not well constrained by observational data. With so many free parameters it is possible there are other good fits that we have not found.

\section{Conclusion.}
\label{sec:Conc} 

We have calculated new model atmospheres and synthetic spectra for carbon stars, using the recent ab initio HCN/HNC linelist of \citet{linepaper}. However selectively removing HNC lines from the HCN/HNC linelist and recalculating our models, we have shown that HNC opacity can have a small but significant effect upon the structure of carbon star atmospheres. The general shape of the pressure temperature curves of our model atmospheres which exclude HNC are consistent with those of \citet{Aoki1} and \citet{Jorgensen2000}. However our models with both HCN and HNC will differ from those of \citet{Aoki1} and \citet{Jorgensen2000}, especially at low effective temperatures (\teff$<2900$ K) and low \CO (\CO$<0.01$).

We have obtained good fits of our synthetic spectra to the observed spectra of WZ Cas and TX Psc. These fits have provided a new independent estimate of the C/O ratio and \Teff\  of these stars. 
In general our estimate of the \Teff\  and C/O ratio for WZ~Cas and TX~Psc are in agreement with earlier works. For WZ~Cas our \Teff\  and C/O ratio are both slightly lower than earlier estimates, this is consistent with an increase in overall opacity within the model due to the inclusion of HNC. For TX Psc our C/O ratio is toward the lower end of the earlier estimates of the C/O ratio. Our models using values of \CO\  greater than 0.02 favoured by earlier works cannot reproduce the observed spectra of TX~Psc. 
However by changing the many free parameters such as log g \CO, \NO\  and \Teff\  it may be possible to obtain other good fits, for example \citet{Jorgensen2000} obtained two good fits with different values of \CO, \Teff\  and log g.

By comparing our synthetic spectra calculated with and without HNC opacity, we have identified the contribution of HNC at 2.9 \mim\  to the spectrum of WZ Cas for the first time. Absorption resulting from the Q branch lines of the HCN $\Delta v_2=1, \Delta v_3=1$ and $\Delta v_1=1, \Delta v_2=-1$ bands has been identified in the spectrum of WZ Cas.
It has been shown that to correctly model the atmospheres of C-stars, opacity arising from HNC must be accounted for.

It may be possible to correct for systematic errors in line frequencies of the linelist of \citet{linepaper} by substituting experimentally determined energy levels for their ab initio counterparts. This would give very accurate line frequencies for the lines corresponding to transitions between the experimentally determined energy levels. 
 Similarly by substituting experimentally determined H$^{13}$CN energy levels for their ab initio H$^{12}$CN counterparts, it should be possible to better account for the spectrum of H$^{13}$CN. This, however, must await the assignment of approximate quantum numbers to the energy levels of the linelist of \citet{linepaper}.

\section*{Acknowledgments}

Our referee Wako Aoki is thanked for helpful comments, which have served to improve this paper. We thank the UK particle physics and astronomy research council (PPARC) for post doctoral funding for GJH. PPARC and the Royal Society are thanked for travel grants. 
The work of YP is partially supported by a research grant from the American Astronomical Society.
The manipulation and analysis of the HCN/HNC linelist was carried out on the Miracle 24-processor Origin 2000 supercomputer at the HiPerSPACE computing centre, UCL, which is part funded by PPARC.

\end{document}